\documentclass[10pt,a4paper]{article}
\usepackage[margin=1in]{geometry}
\usepackage{bm}
\usepackage{latexsym}
\usepackage{dcolumn}
\usepackage{amsmath,amsfonts,amssymb}

\usepackage[font=small,labelfont=bf]{caption}
\usepackage[numbers,square,sort&compress]{natbib}
\usepackage{authblk}
\usepackage{graphicx, amsmath}
\usepackage[latin1]{inputenc}
\usepackage{lineno}

\makeatletter \renewcommand{\@biblabel}[1]{#1.} \makeatother
\graphicspath{%
{figs/}%
}

\begin{document}
\title{Inflation driven by Barrow Holographic dark energy}

\author[1$^\dagger$]{Sayani Maity} 
\author[2$^*$]{Prabir Rudra}

\affil[1]{Department of Mathematics, Techno Main Salt Lake, Sector-V, Kolkata-700091, India}
\affil[2]{Department of Mathematics, Asutosh College, Kolkata-700026, India}

\maketitle

\textit{~~~~~~~~~~~~~~~~~~~~~~~~We dedicate the paper to the memory of John D. Barrow.}

\begin{abstract}
In this work we have investigated the inflation mechanism driven by the Barrow Holographic dark energy (BHDE) in the early universe. BHDE is based on the Barrow relation for horizon entropy, which in turn is inspired from the shape of the COVID-19 virus. It was shown by Barrow that the quantum gravitational effects may instigate complex fractal features in the structure of a black hole. Since the length scale during the inflation is expected to be small, the energy density obtained from the application of the holographic principle in the early universe will be large enough to support the inflationary scenario. Using the Granda-Oliveros IR cut-off we have studied the inflationary scenario with the universe filled with BHDE. Various analytic solutions for the model were found out including the slow-roll parameters, scalar spectral index and tensor-to-scalar ratio. Since inflation is generally attributed to the presence of scalar fields, we have explored a correspondence between BHDE and scalar field models. Both canonical scalar field and the Tachyonic scalar field have been considered for this purpose. The evolution of the potential generated from the fields are plotted and found to be consistent with the observations. From the work we see that BHDE can be a model of dark energy that can successfully drive the early time inflation. 
\end{abstract}


\footnote{$^\dagger$sayani.maity88@gmail.com;
$^*$prudra.math@gmail.com, rudra@associates.iucaa.in}

\section{Introduction}
The inflationary phase is a crucial era in the cosmological evolution of the universe. It is a very short lived phase in the early times just after big bang when the universe experienced extreme expansion. The concept was proposed way back in the 1980s \cite{inf1, inf2, inf3, inf4, inf5} and since then it has received a lot of attention from the researchers. Early inflationary phase along with the late time cosmic acceleration \cite{acc1, acc2} forms the two pillars of the present cosmological model. Since regular matter cannot account for these extremely expanding phases, exotic forms of matter have been proposed. While the late time acceleration has been attributed to the presence of dark energy (DE), the inflationary phase is assumed to be driven by a special form of DE known as the scalar fields with slow rolling assumptions \cite{sl1, sl2, sl3}. Considerable amount of research has been done on inflationary models over the period of last three decades \cite{inf6, inf7, inf8, inf9, inf10, inf11, inf12, inf13, inf14, inf15}.

Although the concept of DE has been able to explain the accelerated expansion of the universe at late times, yet it is universally accepted that the exact nature of DE is not yet known to us. For an extensive review on dark energy the reader is referred to \cite{de1}. There are however different candidates which effectively play the role of DE. Chaplygin gas models are well known candidates of DE \cite{chap1, chap2, chap3, chap4, chap5, chap6}. Similarly scalar fields are also effective DE models that are believed to play important role in inflation. Another potential candidate for DE is the Holographic dark energy (HDE) \cite{hde1, hde2, hde3}, whose formulation is based on the holographic principle \cite{hp1}. The holographic principle, which has its origin in the black hole (BH) thermodynamics, states that the entropy of a system is characterized by its area and not by its volume \cite{holostring, vol1}. The idea was inspired from the insight that the information of all the objects that have fallen into a BH is encoded on the surface of the event horizon. The holographic principle was extended to string theory by Susskind in Ref.\cite{holostring}. Also the de-Sitter space holography is recently formulated by him in Ref.\cite{new1}.

The role of HDE as a dark energy driving the late time cosmic acceleration has been extensively discussed in the literature \cite{hde4, hde5, hde6}. This directly provides motivation to study the role of HDE in the inflationary phase. It is known that the HDE density is related to the inverse square of the infrared (IR) cut-off. Moreover since the IR cut-off is related to causality, it is generally considered as a form of horizon such as the particle horizon, future event horizon or Hubble length. There can be different motivated forms of IR cut-off that can be used in the study of HDE. One such form is the Granda-Oliveros (GO) cut-off which was framed from dimensional motivations \cite{go1, go2} and is given by a combination of the Hubble parameter and its time derivative. Since the length scale during the inflation is expected to be small, the energy density obtained from the application of the holographic principle in the early universe will be large enough to support the inflationary scenario. 

The standard entropy of a physical system is given by $S=A/4$ (where $A$ is the area), and the HDE is based on this relation. But any modification to the entropy-area relation will in principle result in a modified form of HDE. These corrections to the area law are basically inspired from quantum gravitational framework. Loop quantum gravity introduces logarithmic corrections \cite{log1, log2, log3, log4} to the area law, whereas power law corrections are introduced by the entanglements of the quantum fields \cite{pow1, pow2}. Another important consideration was the modification of the thermodynamic entropy of a system to the non additive form instead of the additive one \cite{add1, add2, add3}. Based on this concept it was shown in \cite{newarea} that the standard area law must be generalized to the power law  form $S=\gamma A^{\delta}$, where $\delta$ is the Tsallis parameter, $\gamma$ is an unknown parameter and $A$ is the area of the BH horizon. It can be clearly seen that for $\gamma=1/4$ and $\delta=1$, the above entropy reduces to the standard Boltzmann-Gibbs entropy discussed before. The Tsallis HDE is inspired from this Tsallis entropy and has been investigated widely in literature for the late time acceleration using various IR cut-offs \cite{tsallis1, tsallis2, tsallis3, tsallis4, tsallis5}. The Tsallis HDE as a source of early time inflation was studied in \cite{tsallinf} using the GO infrared cut-off, whereas inflation driven by standard HDE for different IR cut-offs were studied in \cite{standinf1, standinf2}.

Recently, inspired from the shape of the COVID-19 virus, Barrow showed that the quantum gravitational effects may instigate complex fractal features in the structure of a BH. A three-dimensional spherical analogue of a 'Koch Snowflake' was created by an infinite diminishing hierarchy of touching spheres around the Schwarzschild event horizon. Starting with a Schwarzschild BH with mass $M$ and radius $R_g=2GM/c^2$, we keep on attaching some smaller spheres which touch its outer surface, just like way how the COVID-19 virus gets attached to the parent cell. Further smaller spheres are attached to these outer spheres and the sequence is continued to give a highly complicated fractal structure. Here the boundary will be composed of surfaces of hierarchically smaller spheres in contact with each other. This will be like a 'sphereflake' and an animation showing its construction may be visualized in Ref.\cite{movie}. This complicated structure induces a finite volume, but with infinite or finite area, which in turn leads to the modification of the BH entropy \cite{barrow1}
\begin{equation}\label{barrowentropy}
S_{B}=\left(\frac{A}{A_0}\right)^{1+\Delta/2},
\end{equation}
where $A$ is the standard horizon area and $A_0$ is the Planck area. Here the deformation induced by quantum gravity is given by the exponent $\Delta$. For $\Delta=0$, we recover the standard Bekenstein-Hawking entropy with simple structure and for $\Delta=1$ we get the complex fractal structure of the horizon. This toy model is basically intended to show that near the scale where the quantum gravity effects dominates, the surface area of a BH can greatly exceed $4\pi R_{g}^{2}$ due to the presence of intricate fractal features. Moreover this will happen for any external intricacy having a Hausdorff dimension greater than $2$. Basically the 2-dimensional geometrical surface behaves like more than 2 dimensions and tends towards the behaviour of a 3-dimensional surface in a limiting scenario with maximum intricacies. This shows that the 2-dimensional surface behaves as if it has all the information of a 3-dimensional volume. It must be noted that this fractal nature does not arise from any particular quantum gravity calculations, but from generalised physical principles and hence is a reasonable proposition as an initial approach \cite{barrow1}. On larger scales this entropy model has become one of the prototypes for observational analysis. Using this Barrow relation for the horizon entropy Saridakis in \cite{bhde1} constructed the Barrow Holographic dark energy (BHDE). BHDE is a generalized model possessing the usual HDE as a $\Delta=0$ limit. In \cite{bhde1} it was shown that BHDE possesses a far richer structure compared to the usual HDE and also a richer cosmological behaviour. The role of BHDE in the late time acceleration was discussed in detail. Here we are interested in studying the inflationary scenario of the universe driven by BHDE. The paper is organized as follows: In section 2 we discuss the basic equations of BHDE. Section 3 is devoted to the study of inflation driven by BHDE. In section 4, a correspondence between BHDE and scalar field models are set up. Finally the paper ends with some discussion and conclusion in section 5.

\section{Barrow Holographic dark energy}

Here we will construct the BHDE from the Barrow entropy which is a modification to the standard entropy relation. The standard entropy relation is characterized by the inequality $\rho_{DE}L^{4}\leq S$, where the condition $S\propto A\propto L^{2}$ is imposed, $L$ being the horizon length.
The energy density of BHDE as motivated from the Barrow entropy relation (\ref{barrowentropy}) is given by
\begin{equation}\label{2.1}
    \rho_{BHDE}=C L^{\Delta-2},\mbox{where}~~C=3c^{2} M_P^{2}.
\end{equation}
$\Delta$ is the parameter that measures the quantum gravitational deformation, $M_P$ is the Planck mass, $c^2$ is the model parameter and $L$ is the IR cut-off length. For $\Delta=0$, eqn.(\ref{2.1}) reduces to the energy density for the standard holographic dark energy, and as $\Delta \rightarrow 1$ it becomes modified and expresses the fractal features. Here it should be stated that the parameter $C$ has the dimensions $[L]^{-\Delta-2}$.

Here we consider the flat Friedmann-Lemaitre-Robertson-Walker (FLRW) metric of the form,
\begin{equation}
ds^{2}=-dt^{2}+a(t)^{2}\left(dx^{2}+dy^{2}+dz^{2}\right),
\end{equation}
where $a(t)$ is the cosmological scale factor representing the expansion of the universe. The Hubble parameter $H$ is defined as,
\begin{equation}
H\equiv \frac{\dot{a}}{a},
\end{equation}
where the dot $(.)$ derivative with respect to time.

The Friedmann equations in a homogeneous and flat FLRW
universe filled with the dark energy and matter read as
 \begin{equation}\label{2.2}
     H^2=\frac{1}{3M_P^{2}}(\rho_{BHDE}+\rho_m),
 \end{equation}
 and
 \begin{equation}\label{2.3}
 2 \Dot{H}+3 H^{2}=-\frac{1}{M_P^{2}} p_{BHDE},
 \end{equation}
 where $\rho_{m}$ and $p_{BHDE}$ represent density of matter and pressure of BHDE respectively.
 Assuming that there is no interaction between dark energy and matter, we have the conservation equation for BHDE as
 \begin{equation}\label{2.4}
 \Dot{\rho}_{BHDE}+3H(1+\omega_{BHDE})\rho_{BHDE}=0,
 \end{equation}
 where $\omega_{BHDE}=p_{BHDE}/\rho_{BHDE}$ is the equation of state (EoS) parameter corresponding to BHDE. Exploiting Friedmann equations  $\omega_{BHDE}$ reads
 \begin{equation}\label{2.5}
     \omega_{BHDE}=-1-\frac{2M_P^{2}\Dot{H}}{CL^{\Delta-2}},
 \end{equation}
 The most simple choice for the IR cut-off $L$ is the inverse Hubble length $H^{-1}$. But there other choices such as the particle horizon or the future horizon, etc. GO cut-off is one such alternative as already discussed before. Here the length scale is taken as a combination of the Hubble parameter and its time gradient as given below,
 \begin{equation}\label{2.6}
 L^{-2}=\alpha H^2+\beta \Dot{H},
 \end{equation}
 where $\alpha$ and $\beta$ are dimensionless parameters. Since the entropy already includes the quantum gravitational corrections, it is not required to introduce high energy regime modifications to the GO cut-off.

\section{Inflation driven by Barrow Holographic dark energy}
 We have assumed that the inflationary era of the Universe is driven by the BHDE with the IR cut-off as GO cut-off given by (\ref{2.6}). Ignoring the contribution of matter part, the Friedmann equation for an expanding Universe reads
\begin{equation}\label{3.1}
    H^2=\frac{1}{3M_P^{2}} C(\alpha H^2+\beta \Dot{H})^{1-\frac{\Delta}{2}},
\end{equation}
Exploiting equations (\ref{2.1}) and (\ref{2.6}) we get
\begin{equation}\label{3.2}
    \Dot{H}=\frac{H^2}{\beta}\left[ \left(\frac{3M_P^{2}}{C}\right)^{\frac{2}{2-\Delta}}(H^{2})^{\frac{\Delta}{2-\Delta}} -\alpha \right].
\end{equation}
For change of variable we consider $N=\ln (a/a_{i})$, where $a_{i}$ is initial value of $a$. From this variable change we have $dN=Hdt$ and $\Dot{H}=\frac{1}{2}\frac{d H^{2}}{dN}$. Integrating (\ref{3.2}) we obtain Hubble parameter in terms of number of e-folds as
\begin{equation}\label{3.3}
    \ln \left[(\widetilde{H}^{2})^{\frac{\Delta}{2-\Delta}} -\left(\frac{3M_P^{2}}{C}\right)^{\frac{2}{2-\Delta}}\left(M_P^{2}\right)^{\frac{\Delta}{2-\Delta}}\frac{1}{\alpha}\right]_{\widetilde{H_{i}}}^{\widetilde{H_{e}}}= \ln C_{1}+\frac{\Delta}{2-\Delta} \ln (M_P^{2})-\frac{2\alpha \Delta N}{\beta (\Delta-2)},
\end{equation}
where $\widetilde{H}=H/M_P$ is the dimensionless Hubble parameter. Exploiting equation (\ref{3.2}) the first slow-roll parameter is derived as
\begin{equation}\label{3.4}
    \epsilon_{1}=-\frac{\Dot{H}}{H^{2}}=-\frac{1}{\beta}\left[\left(\frac{3M_P^{2}}{C}\right)^{\frac{2}{2-\Delta}}\left(M_P^{2}\right)^{\frac{\Delta}{2-\Delta}}(\widetilde{H}^{2})^{\frac{\Delta}{2-\Delta}}-\alpha\right],
\end{equation}
The other slow-roll parameters are defined as, \cite{1KDI95,1SMSED96,1MJRCVV14,1WRP14}
\begin{equation}\label{3.5}
    \epsilon_{n+1}=\frac{d\ln (\epsilon_{n}) }{dN},
\end{equation}
Using the above iterative relation the second slow-roll parameter can be derived as
\begin{equation}\label{3.6}
    \epsilon_{2}=\frac{\dot{\epsilon_{1}}}{H\epsilon_{1}}=\frac{2}{\beta}\left(\frac{3M_P^{2}}{C}\right)^{\frac{2}{2-\Delta}}\left(\frac{\Delta}{2-\Delta} \right)\left(M_P^{2}\right)^{\frac{\Delta}{2-\Delta}}(\widetilde{H}^{2})^{\frac{\Delta}{2-\Delta}},
\end{equation}
These slow-roll parameters are considered to be very small at the start of the inflationary phase. At $\epsilon_{1}=1$ the inflationary phase ends. The Hubble parameter at this epoch takes the form
\begin{equation}\label{3.7}
    \widetilde{H_{e}}^{2}=\left[(\alpha-\beta)\left(\frac{C}{3M_P^{2}}\right)^{\frac{2}{2-\Delta}} \left(M_P^{2}\right)^{\frac{\Delta}{\Delta-2}}\right]^{\frac{2-\Delta}{\Delta}}.
\end{equation}

Using equation (\ref{3.3}) the Hubble parameter is derived at the earlier phase of inflation that includes the horizon crossing time as
\begin{equation}\label{3.8}
    \widetilde{H_{i}}^{2}=\frac{1}{M_P^{2}}\left[\frac{1}{\alpha-\beta} \left(\frac{3M_P^{2}}{C}\right)^{\frac{2}{2-\Delta}}-C_{1} e^{-\frac{2\alpha \Delta N}{\beta (\Delta-2)}}\right]^{\frac{\Delta-2}{\Delta}}.
\end{equation}
According to ref \cite{1MJRCVV14}, the expressions of 
the inflationary observables, namely the scalar
spectral index of the curvature perturbations $n_{s}$ and the tensor-to-scalar ratio $r$ are given by
\begin{equation}\label{3.9}
    r= 16\epsilon_{1} ~~~~~ \mbox{and}~~~~n_{s}=1-2\epsilon_{1}-2\epsilon_{2}.
\end{equation}
The above expressions of the observables are approximate values, since they have not been obtained from a complete perturbation analysis of holographic dark energy. Hence it is obvious that a full perturbation analysis is required to extract the exact expressions for the inflationary observable. But here we have adopted a procedure that also provides good approximation as long as $H(t)$ is known. For this we have calculated these parameters at the time of horizon crossing that result in
 \begin{equation}\label{3.10}
     r\mid_{\widetilde{H_{i}}}=-\frac{16}{\beta}\left[\left \{ \frac{1}{\alpha-\beta}-C_{1}e^{-\frac{2\alpha \Delta N}{\beta (\Delta-2)}} \left(\frac{C}{3M_P^{2}}\right)^{\frac{2}{2-\Delta}}\right \}^{-1}-\alpha \right],
 \end{equation}
 and
  \begin{eqnarray*}
     n_{s}\mid_{\widetilde{H_{i}}}=1+\frac{2}{\beta}\left[\left \{ \frac{1}{\alpha-\beta}-C_{1}e^{-\frac{2\alpha \Delta N}{\beta (\Delta-2)}} \left(\frac{C}{3M_P^{2}}\right)^{\frac{2}{2-\Delta}}\right \}^{-1}-\alpha \right]+\frac{4}{\beta}\left(\frac{3M_P^{2}}{C}\right)^{\frac{2}{2-\Delta}}\left(\frac{\Delta}{2-\Delta} \right)
 \end{eqnarray*}
  \begin{equation}\label{3.11}
  \left[ \frac{1}{\alpha-\beta}\left(\frac{3M_P^{2}}{C}\right)^{\frac{2}{2-\Delta}}-C_{1}e^{-\frac{2\alpha \Delta N}{\beta (\Delta-2)}}\right]^{-1},
  \end{equation}
  
  These approximate expressions are quite efficient in describing the scenario and can be consistently used with the observational data to constrain the parameter space of the model.
  
\section{Correspondence between BHDE and scalar field}
This section is dedicated to investigate that if it is possible to describe the inflationary paradigm by BHDE approach through the dynamics of a scalar field. Here we consider two different models namely the Canonical scalar field and the Tachyonic field.
\subsection{Canonical Scalar field }
Firstly the correspondence between BHDE
and canonical scalar field is considered. According to Ref. \cite{1BKCSNSOSD12}, the energy density and pressure of canonical scalar field read
\begin{equation}\label{4.1}
\rho_{\phi_c}=\frac{1}{2}\dot{\phi_c}^{2}+V(\phi_c),
\end{equation}
$\mbox{and}$
\begin{equation}\label{4.1.1}
p_{\phi_c}=\frac{1}{2}\dot{\phi_c}^{2}-V(\phi_c). 
\end{equation}
where $\phi_c$ represents the canonical scalar field.
Comparing the energy density and pressure of BHDE and canonical scalar field we get
\begin{equation}\label{4.2}
V(\phi_c)=\rho_{BHDE}-\frac{1}{2}\dot{\phi_c}^{2}=\frac{1}{2}\rho_{BHDE}(1-\omega_{BHDE}),
\end{equation}
$\mbox{and}$
\begin{equation}\label{4.3}
\Dot{\phi_c}^{2}=\rho_{BHDE}+p_{BHDE}=\rho_{BHDE}(1+\omega_{BHDE}).
\end{equation}
Inserting (\ref{2.5}) and (\ref{3.2}) in (\ref{4.2}) we have obtained the potential in terms of Hubble parameter as
\begin{equation}\label{4.4}
V(\phi_c)=M_P^{4}\widetilde{H}^{2}\left[ 3+\frac{1}{\beta}\left \{\left(\frac{3M_P^{2}}{C}\right)^{\frac{2}{2-\Delta}} \left(M_P^{2}\widetilde{H}^{2}\right)^{\frac{\Delta}{2-\Delta}}-\alpha \right \} \right],
\end{equation}

The constructed potential function from the canonical scalar field is plotted in Fig.1 against the Hubble parameter $H$. Different trajectories are generated for different values of $\Delta$ to understand the dependence of the potential on the parameter. From the figure it is evident that the potential grows with the Hubble function. This result is consistent with the observations.

\vspace{6mm}

\begin{figure}[hbt!]
\centering
\includegraphics{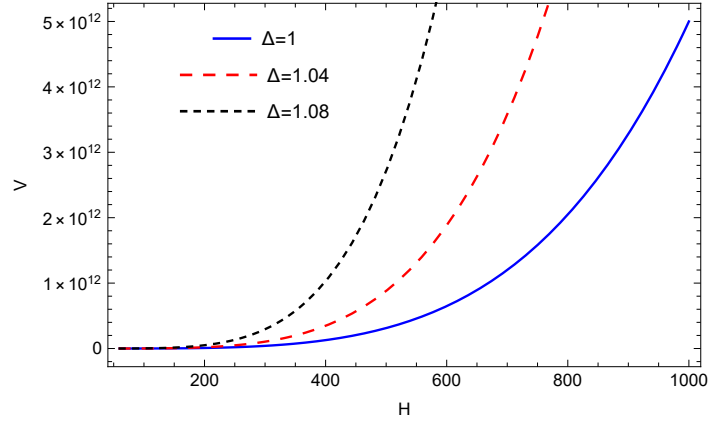}
\centering \caption{The constructed potential $V(\phi_c)$ for BHDE from the canonical scalar field is plotted against the Hubble parameter $H$. The constants are considered as $M_{p}=1$, $C=3$, $\alpha=0.1$, $\beta=0.2$.}
\label{Fig1a}
\end{figure}

Now the expression of $\Dot{\phi_c}^{2}$ can be obtained from Friedmann equation as $\Dot{\phi_c}^{2}=-2M_P^{2} \Dot{H}$. For a change of variable it can be rewritten as $\Dot{\phi_c}=H \phi_c'$ and 
\begin{equation}\label{4.3.1}
\phi_c'^{2}=-2M_P^{2} \frac{\Dot{H}}{H^{2}},
\end{equation} 
where $\phi_c'$ represents derivative of $\phi_c$ with respect to number of e-folds $N$, i.e. $\phi_c'=d\phi_c/dN$. Exploiting equation (\ref{3.4}) and integrating (\ref{4.3.1}) one arrives at the following expression
\begin{equation}\label{4.5}
   \Delta\phi_c N=\sqrt{2} M_P \int_0 ^N\sqrt{-\frac{1}{\beta}\left \{\left(\frac{3M_P^{2}}{C}\right)^{\frac{2}{2-\Delta}} \left(M_P^{2}\widetilde{H}^{2}\right)^{\frac{\Delta}{\Delta-2}}-\alpha \right \}}dN,
\end{equation}
which can be solved numerically (since it is difficult to obtain an analytic solution with the known mathematical rules) to obtain the evolution of the potential function with respect to the field $\phi_c$. Here $N=0$ represents the horizon crossing of perturbation.

\subsection{Tachyonic Field}

This section aims to investigate the condition for which BHDE behaves like a Tachyonic field. The energy density and pressure of the Tachyonic field \cite{1SA02} are as follows

\begin{equation}\label{4.6.1}
\rho_{\phi_T}=\frac{V(\phi_T)}{\sqrt{1-\Dot{\phi_T}^{2}}},
\end{equation}
\mbox{and}
\begin{equation}\label{4.6.2}
p_{\phi_T}=-V(\phi_T) \sqrt{1-\Dot{\phi_T}^{2}}.
\end{equation}
where $\phi_T$ represents the tachyonic scalar field.
To determine an appropriate potential for tachyonic field for which it behaves like BHDE, we compare the energy densities and pressure of these two dark energy models, which results in
\begin{equation}\label{4.7}
\Dot{\phi_T}^{2}=1+\omega_{BHDE}.
\end{equation}
\mbox{and}
\begin{equation}\label{4.8}
V(\phi_T)=\rho_{BHDE}\sqrt{1-\Dot{\phi_T}^{2}}.
\end{equation}
Using (\ref{2.5}) and (\ref{3.4}) we have
\begin{equation}\label{4.9}
\acute{\phi_T}^{2}=\frac{2}{3}\epsilon_{1}\left(M_P^{2}\widetilde{H}^{2}\right)^{1-\frac{\Delta}{2}}.
\end{equation}
Integration of eqn.(\ref{4.9}) results in
\begin{equation}\label{4.10}
\Delta\phi_{T} N=\sqrt{\frac{2}{3}\left(M_P^{2}\right)^{1-\frac{\Delta}{2}}} \int_0 ^N \sqrt{-\frac{\widetilde{H}^{2}}{\beta}\left[\left(\frac{3M_P^{2}}{C}\right)^{\frac{2}{2-\Delta}}\left(M_P^{2}\right)^{\frac{\Delta}{2-\Delta}}(\widetilde{H}^{2})^{\frac{\Delta}{2-\Delta}}-\alpha\right]}dN.
\end{equation}

The constructed potential function from the tachyonic scalar field is plotted in Fig.2 against the Hubble parameter $H$. Here also different trajectories are generated for different values of $\Delta$ to understand the dependence of the potential on the parameter. From the figure it is evident that the potential grows with the Hubble function. This result is consistent with the observations.

\begin{figure}[hbt!]
\centering
\includegraphics{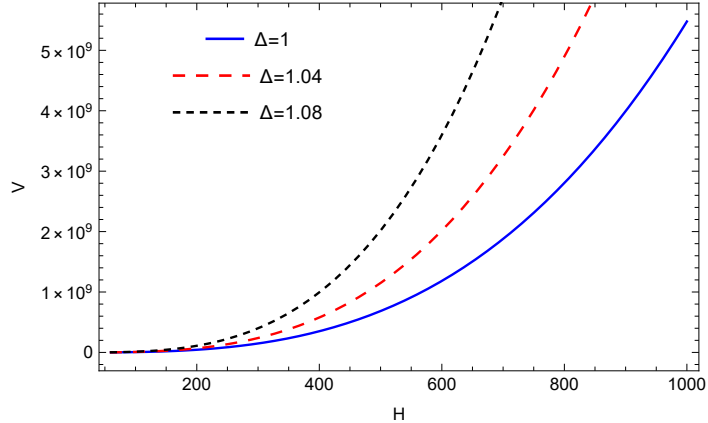}
\centering \caption{The constructed potential $V(\phi_T)$ for BHDE from the Tachyonic scalar field is plotted against the Hubble parameter $H$. The constants are considered as $M_{p}=1$, $C=3$, $\alpha=0.1$, $\beta=0.2$.}
\label{Fig2a}
\end{figure}

\section{Discussion and Conclusion}
The application of the holographic principle in the late time acceleration is quite common in literature. Motivated from this, here we have applied the principle in the early time inflationary scenario. According to the principle, the energy density is proportional to the inverse of length squared. Since the length scale is expected to be very small during inflation, the energy density generated should be large enough to drive the inflation. It is known that the holographic dark energy originates from the entropy, the nature of the DE can be altered by modifying the entropy law. Such modifications can be brought about by considering quantum gravitational effects. Here we have considered one such modification by considering the entropy relation of Barrow, which was motivated from the shape of the COVID-19 virus. It was found that these quantum corrections induced complex fractal features in the structure of a black hole. We have explored an inflationary scenario described by a universe filled with Barrow holographic dark energy, using the Granda-Oliveros IR cut-off. Various analytic solutions for the model were found out including the slow-roll parameters, scalar spectral index and tensor-to-scalar ratio. Finally a correspondence between the BHDE and scalar fields is explored. Both canonical scalar field and Tachyonic field is used for this purpose. The potential generated from the two different fields are plotted to get an idea of its evolution. It is seen that the trend is consistent with the observational evidences. This work shows that BHDE can be a perfect candidate to drive the early inflationary scenario of the universe. This along with its success in explaining the late acceleration makes the model a successful candidate of dark energy that can satisfactorily explain almost all the major phases of cosmological evolution of the universe.

\section*{Acknowledgments}

P.R. acknowledges the Inter University Centre for Astronomy and
Astrophysics (IUCAA), Pune, India for granting visiting
associateship.

\end{document}